\documentclass[twocolumn,showpacs,preprintnumbers,amsmath,amssymb]{revtex4}

\usepackage[dvips]{graphicx}

\usepackage{epsfig}
\usepackage{soul}
\begin{document}

\title{Manipulating sudden death of entanglement of two-qubit X-states in thermal reservoirs}

\author{Mazhar Ali$^1$, G.\ Alber$^1$ and A.\ R.\  P. Rau$^2$}
\affiliation{$^1$Institut f\"{u}r Angewandte Physik, Technische Universit\"{a}t Darmstadt, D-64289, Germany \\
$^2$Department of Physics and Astronomy, Louisiana State University,
Baton Rouge, Louisiana 70803, USA}

\begin{abstract}
Manipulation of sudden death of entanglement (ESD) of two qubits interacting with statistically uncorrelated thermal reservoirs is investigated. It is shown that for initially prepared X-states of the two qubits a simple (necessary and sufficient) criterion for ESD can be derived with the help of the Peres-Horodecki criterion. It is shown analytically that, in contrast to the zero-temperature case, at finite temperature of at least one of the reservoirs all initially prepared two-qubit X-states exhibit ESD. General conditions are derived under which ESD can be hastened, delayed, or averted.
\end{abstract}


\maketitle

\section{Introduction}

Entanglement is a vital resource for quantum information processing \cite{Nielsen2000}. Numerous processes relevant for quantum computation, quantum cryptography \cite{E-PRL91} or quantum teleportation \cite{BBCJPW-PRL93} rely on entangled qubit states. Recently, Yu and Eberly \cite{YE-PRB03,YE-PRL04,YE-OC06,YE-PRL06,EY-Sc07} found that reservoir-induced decay of single-qubit coherence can be slower than the corresponding decay of qubit entanglement. Both, abrupt and asymptotically gradual disappearance of entanglement were predicted in amplitude \cite{YE-PRL04} and phase damping channels \cite{YE-OC06}. Abrupt disappearance of entanglement in finite time was termed ``entanglement sudden death" (ESD). Also recently, experimental evidence of ESD was reported for an optical setup \cite{AM-Sc07} and atomic ensembles \cite{LC-PRL07}. Whereas first investigations on ESD concentrated on entangled two-qubit states, later it was also explored in a wider context and in higher dimensional Hilbert spaces \cite{LR-PRA07,SW-PRA07,AQJ-PRA08,ILZ-PRA07,AJ-PLA07,ARR-Arx07,AJ-PRA07,CZ-PRA08,DJ-PRA06}.

Clearly, ESD is a serious limiting factor for the use of entangled qubits in quantum information processing. It would seem important to stabilize quantum systems against this unwanted phenomenon. First studies in this context concentrated on changing initial states to more robust ones of the same degree of entanglement \cite{J-JPA06, YE-QIQC07}. Recently, we have addressed the practically relevant question whether it is possible to delay or even avert ESD by application of particularly chosen local unitary transformations for a given initial state and a given open-system dynamics \cite{RAA-Arx07}. We demonstrated that this is indeed possible for the special two-qubit system investigated first by Yu and Eberly \cite{YE-PRL04} and have found similar effects in qubit-qutrit systems \cite{ARA-UP08}.

In this paper, we generalize our previous results and investigate ESD of two qubits which are interacting with statistically independent (bosonic) reservoirs at finite temperatures. It is demonstrated that based on the Peres-Horodecki criterion \cite{Pe-PRL96,HHH-PLA96}
and on recent results of Huang and Zhu \cite{HZ-PRA07} it is possible to develop systematically a simple criterion capable of characterizing delay and
avoidance of ESD of initially prepared two-qubit X-states in this open quantum system. With the help of this criterion it is proved that, in agreement with recent conjectures based on numerical case studies \cite{ILZ-PRA07}, all initially prepared two-qubit X-states exhibit ESD if at least one of the statistically independent reservoirs is at finite temperature. However, if both reservoirs are at zero temperature, there are always some X-states for which ESD does not occur. Based on this criterion, we demonstrate that even at finite temperatures of the reservoirs it is possible to hasten or to delay ESD. The characteristic time of ESD can be controlled by the time when appropriate local unitary operations are applied. However, unlike in the zero-temperature case, when at least one of the reservoirs is at finite temperature it is not possible to avoid ESD completely by any choice of local unitary transformations. 

\section{Open-system dynamics of two qubits coupled to statistically independent thermal reservoirs\label{II}}

In this section, we briefly summarize the basic equations of motion governing our open quantum system of interest. In order to put the problem of ESD, its delay, and avoidance into perspective, let us consider two non-interacting qubits which are spatially well separated so that each of them interacts with its own thermal reservoir. These two reservoirs are assumed to be statistically independent and are possibly also at different temperatures.
\begin{figure}[h]
\scalebox{1.7}{\includegraphics[width=1.3in]{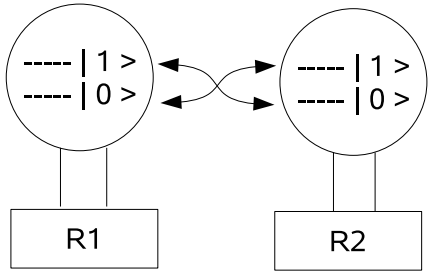}}
\caption{Schematic representation of the two non-interacting two-level atoms (qubits) A and B, initially prepared in an entangled state.
Each of them interacts with its own local reservoir $R_1$ and $R_2$.}
\label{FIG:1}
\end{figure}
The coupling of these two qubits to the reservoirs can originate physically from the coupling of two two-level atoms to the (resonant) modes of the electromagnetic radiation field, for example, with the local radiation field at thermal equilibrium. 

In the interaction picture and in the dipole- and rotating wave approximation, the resulting equation of motion of these two qubits is given by the master equation \cite{Scully-Zubairy1997,ILZ-PRA07}
\begin{eqnarray}
\frac{d \rho}{dt} = \frac{\gamma_1}{2} (m + 1) [2 \sigma_{-}^1 \rho \sigma_{+}^1 - \sigma_{+}^1 \sigma_{-}^1 \rho - \rho \sigma_{+}^1 \sigma_{-}^1] \nonumber \\ + \frac{\gamma_1}{2} m [2 \sigma_{+}^1 \rho \sigma_{-}^1 - \sigma_{-}^1 \sigma_{+}^1 \rho - \rho \sigma_{-}^1 \sigma_{+}^1] \nonumber \\ + \frac{\gamma_2}{2} (n + 1) [2 \sigma_{-}^2 \rho \sigma_{+}^2 - \sigma_{+}^2 \sigma_{-}^2 \rho - \rho \sigma_{+}^2 \sigma_{-}^2] \nonumber \\ + \frac{\gamma_2}{2} n [2 \sigma_{+}^2 \rho \sigma_{-}^2 - \sigma_{-}^2 \sigma_{+}^2 \rho - \rho \sigma_{-}^2 \sigma_{+}^2].
\label{eq:1}
\end{eqnarray}
For reservoirs representing the electromagnetic radiation field, $m$ and $n$ denote the mean photon numbers of the local reservoirs coupling to qubits $1$ and $2$. The spontaneous emission of atom $i$ from its excited state $|1_i\rangle$ to its ground state $|0_i\rangle$ is described by the spontaneous decay rate $\gamma_{i}$ and $\sigma_{\pm}^i$ are the corresponding raising (+) and
lowering (-) operators, i.e. $\sigma_{+}^i = |1_i\rangle\langle0_i|$ and $\sigma_{-}^i = |0_i\rangle\langle1_i|$. The orthonormal atomic eigenstates 
$|1\rangle = |1 \rangle_A \otimes |1\rangle_B$,
$|2\rangle = |1\rangle_A \otimes |0\rangle_B$,
$|3\rangle = |0\rangle_A \otimes |1\rangle_B$,
$|4\rangle = |0\rangle_A \otimes |0\rangle_B$ 
form the (computational) basis of the four dimensional Hilbert space of the two qubits. The derivation of Eq.(\ref{eq:1}) also assumes the validity of the Born-Markov approximation. The general solution valid for an arbitrary initially prepared two-qubit state is given in Appendix A by Eq.(\ref{A1}). 

There is some numerical evidence \cite{ILZ-PRA07} that the presence of non-zero mean thermal photon numbers in Eq.(\ref{eq:1}) may be responsible for ESD. However, systematic exploration of these phenomena which is capable of proving the sufficiency of non-zero photon numbers for ESD and of providing  a systematic analytical understanding of ESD in zero-temperature reservoirs has been missing so far. Our main purpose is to close this gap. In particular, we shall develop a simple analytical criterion for ESD which allows a systematic understanding of ESD and its delay and avoidance for initially prepared entangled X-states.

\section{The Peres-Horodecki criterion and entanglement sudden death}

In this section, we analyze conditions under which initially prepared entangled two-qubit states evolving according to Eq.(\ref{eq:1}) exhibit ESD. Starting from the general time dependent solution of these equations ESD is analyzed with the help of the Peres-Horodecki criterion \cite{Pe-PRL96,HHH-PLA96} and with the help of all principal minors of the partially transposed time-dependent two-qubit quantum state. 

To check for separability of a quantum state, Peres \cite{Pe-PRL96} and Horodeckis \cite{HHH-PLA96} developed a powerful necessary and sufficient condition valid for $ 2 \otimes 2$- and $2 \otimes 3$-systems. It states that in these systems a quantum state is separable if and only if its partially transposed density operator is also a valid quantum state. Based on this observation, a measure of entanglement, called {\it negativity}, was proposed \cite{VW-PRA02}. It is the sum of the absolute values of all the negative eigenvalues of the partially transposed density operator. For $2 \otimes 2$ systems, there can be at most one such negative eigenvalue \cite{STV-PRA98}. 

Recently, Huang and Zhu \cite{HZ-PRA07} studied the Peres-Horodecki criterion by focussing on the principal minors of the partially transposed density matrix. The principal minor $[\rho^{PT}(ijkl...)]$ of the partially transposed density operator $\rho^{PT}$ is the determinant of the submatrix ${\cal M}(ijkl...)$ formed by the matrix elements of the $i, j, k, l, ...$ rows and columns of the partial transpose $\rho^{PT}$, that is,
\begin{eqnarray}
{\cal M}(ijkl...) &=&
\left(
\begin{array}{ccccc}
\rho^{PT}_{ii},& \rho^{PT}_{ij},& \rho^{PT}_{ik},& \rho^{PT}_{il},&\cdots\\
\rho^{PT}_{ji},& \rho^{PT}_{jj},& \rho^{PT}_{jk},& \rho^{PT}_{jl},&\cdots\\
\rho^{PT}_{ki},& \rho^{PT}_{kj},& \rho^{PT}_{kk},& \rho^{PT}_{jl},&\cdots\\
\rho^{PT}_{li},& \rho^{PT}_{lj},& \rho^{PT}_{lk},& \rho^{PT}_{ll},&\cdots\\
\cdots&\cdots&\cdots&\cdots&\cdots
\end{array}
\right) .
\label{eq:2}
\end{eqnarray}

In general, if a matrix is positive semidefinite, then all its principal minors are non-negative, and vice versa \cite{HJ-1985}. Therefore, for an entangled two-qubit state $\rho$, the smallest principal minor of its partially transposed density operator must be negative. For two-qubit states the non-negativity of the principal minors $[\rho^{PT} (1)]$, $[\rho^{PT} (2)]$, $[\rho^{PT} (3)]$, $[\rho^{PT} (4)]$, $[\rho^{PT} (12)]$, $[\rho^{PT} (13)]$, $[\rho^{PT} (24)]$, and $[\rho^{PT} (34)]$ is guaranteed already by the non-negativity of the original density matrix $\rho$. As a consequence, a general two-qubit state is entangled if and only if the minimum value of the remaining seven principal minors $P(\rho^{PT})$ is negative, that is,
\begin{widetext}
\begin{equation}
P(\rho^{PT}) \equiv min \{ \, [\rho^{PT}(14)], [\rho^{PT}(23)], [\rho^{PT}(123)], [\rho^{PT}(124)], [\rho^{PT}(134)], [\rho^{PT}(234)], [\rho^{PT}(1234)]\, \} < 0.
\end{equation}
\end{widetext}

Typically, the investigation of these seven principal minors for general solutions $\rho(t)$ of the density operator equation (\ref{eq:1}) is cumbersome. However, significant simplifications are possible for identical zero-temperature reservoirs with $m = n = 0$, and $\gamma_1 = \gamma_2 = \gamma$ in Eq.(\ref{eq:1}). Thus, Huang and Zhu \cite{HZ-PRA07} could demonstrate that for asymptotically long times $t^-_{\infty}$ with $\gamma t^-_{\infty} \gg 1$, the separability of $\rho(t_{\infty}^-)$ is determined by the initial state $\rho$. In particular, these authors showed that the matrix
\begin{eqnarray}
\tilde{\rho} = \left( 
\begin{array}{cccc}
\rho_{11} & \rho_{21} & \rho_{13} & \rho_{23} \\ 
\rho_{12} & \rho_{11} + \rho_{22} & \rho_{14} & \rho_{13} + \rho_{24} \\ 
\rho_{31} & \rho_{41} & \rho_{11} + \rho_{33} & \rho_{21} + \rho_{43} \\
\rho_{32} & \rho_{31} + \rho_{42} & \rho_{12}+ \rho_{34} & 1
\end{array}
\right),
\label{eq:4}
\end{eqnarray}
which is defined in terms of the matrix elements of  the initially prepared two-qubit quantum state $\rho$, determines the asymptotic separability of the two qubits and thus the presence or absence of ESD. 

If $P(\tilde{\rho}) < 0$, the two-qubit state $\rho(t_{\infty}^-)$ is entangled and ESD does not occur. If $P(\tilde{\rho}) > 0$, the asymptotic quantum state $\rho(t_{\infty}^-)$ is separable and ESD takes place. In summary, Huang and Zhu \cite{HZ-PRA07} showed that for identical zero-temperature reservoirs the necessary and sufficient condition for ESD is given by
\begin{eqnarray}
P(\tilde{\rho}) > 0,
\label{eq:5}
\end{eqnarray}
and is determined by the initially prepared two-qubit quantum state.

For initially prepared X-states, that is, quantum states with $\rho_{12}=\rho_{13}=\rho_{24}=\rho_{34}=0$ in the basis of Sec.\ref{II}, this condition can be simplified considerably. In Appendix B, all relevant seven principal minors of a given quantum state $\tilde{\rho}$ 
are evaluated. From these expressions it is apparent that for initially prepared X-states, all the seven relevant principal minors of $\tilde{\rho}$ of Eq.(\ref{eq:4}) are positive if and only if the two principal minors are positive, that is, 
\begin{eqnarray}
[\tilde{\rho}(14)] &,& [\tilde{\rho}(23)]~>~0.
\label{eq:6}
\end{eqnarray}
Thus, in the case of identical zero-temperature reservoirs and for initially prepared X-states, only these two principal minors determine whether the time evolution of Eq.(\ref{eq:1}) causes ESD or not.

Even more general properties can be deduced from the explicit expressions for the principal minors of X-states as given in Appendix B. It is apparent from the general solution of the density operator Eq.(\ref{eq:1}) as given in Appendix A that an initially prepared two-qubit X-state remains an X-state for all times. Combining this observation with the results of Appendix B leads to the general conclusion that initially prepared two-qubit X-states exhibit ESD if and only if at asymptotic times $t_{\infty}^-$ the principal minors
are both positive, i.e.
\begin{eqnarray}
[\rho(14)](t_{\infty}^-)~&,&
[\rho(23)](t_{\infty}^-)~>~0.
\label{eq:7}
\end{eqnarray}
This general result holds even if the two qubits interact with statistically independent reservoirs at finite temperatures so that their dynamics are described by Eq.(\ref{eq:1}).

\section{Two-qubit X-states and quantum control of ESD}

In this section, we specialize our discussion of ESD to initially prepared arbitrary two-qubit X-states. Delay and avoidance of ESD of initially prepared two-qubit X-states coupled to statistically independent zero-temperature reservoirs is discussed in subsection A. In subsection B these results are generalized to reservoirs at finite temperatures. In particular, it is proved that if at least one of the reservoirs has nonzero temperature, all initially prepared X-states exhibit ESD.

Let us first of all briefly summarize some basic properties of X-states. The density matrix of a two-qubit X-state is of the general form
\begin{eqnarray}
\rho_X = \left( 
\begin{array}{cccc}
\rho_{11} & 0 & 0 & \rho_{14} \\ 
0 & \rho_{22} & \rho_{23} & 0 \\ 
0 & \rho_{32} & \rho_{33} & 0 \\
\rho_{41} & 0 & 0 & \rho_{44}
\end{array}
\right).
\label{eq:8}
\end{eqnarray}
i.e. $\rho_{12} = \rho_{13} = \rho_{24} = \rho_{34} = 0$. In particular, Werner states \cite{Wer-PRA89} are special cases of such X-states and some aspects of their ESD have already been discussed  \cite{YE-PRL06,YE-QIQC07,JJ-PLA04,SMD-PRA06}.
Eq.~(\ref{eq:8}) describes a quantum state provided the unit trace and positivity conditions $\sum_{i=1}^4 \rho_{ii} = 1$,  $ \rho_{22} \rho_{33} \geq |\rho_{23}|^2$, and $ \rho_{11} \rho_{44} \geq |\rho_{14}|^2$ are fulfilled. X-states are entangled if and only if either $\rho_{22} \rho_{33} < |\rho_{14}|^2$ or $\rho_{11} \rho_{44} < |\rho_{23}|^2$. Both conditions cannot hold simultaneously \cite{STV-PRA98}.

\subsection{Delaying and avoiding ESD in statistically independent zero-temperature reservoirs}

As discussed in the previous section, for zero-temperature reservoirs the criterion for ESD is given by Eq.(\ref{eq:6}) which together with the results of Appendix B yields the necessary and sufficient conditions for ESD: 
\begin{eqnarray}
[\tilde{\rho}(14)] =&& \rho_{11} - |\rho_{23}|^2~>~0,\nonumber\\ 
&& [\tilde{\rho}(23)] = (\rho_{11} + \rho_{22}) (\rho_{11} + \rho_{33}) - |\rho_{14}|^2 > 0. \nonumber\\
\label{eq:9}
\end{eqnarray}
Depending on the degree of entanglement of the initially prepared two-qubit state, two different cases can be distinguished.

{\bf Case 1:} For initially prepared entangled two-qubit states fulfilling the condition,
\begin{eqnarray}
\rho_{11} \rho_{44} &<& |\rho_{23}|^2, 
\label{eq:10}
\end{eqnarray}
the analytical expression for the negativity of the quantum state $\rho(t)$ satisfying Eq.(\ref{eq:1}) for $\gamma_1=\gamma_2=\gamma$ and $m=n=0$ is given by
\begin{widetext}
\begin{equation}
N_1 (\rho_{X}(t)) = max \, [\, 0, \, \sqrt{F(p,\rho_{ii})^2 - 4 p^2 (\rho_{11} F(p,\rho_{ii}) - p^2 \rho_{11}^2 - |\rho_{23}|^2)} - F(p,\rho_{ii}) \, ],
\end{equation}
\end{widetext}
with $F(p,\rho_{ii}) = (1-2 p + 2 p^2) \rho_{11} + (1-p) (\rho_{22} + \rho_{33}) + \rho_{44}$, $p= {\rm exp}(-\gamma t)$. For any initially entangled two-qubit state $\rho$, Eq.(\ref{eq:10}) implies $\rho_{22} \rho_{33} \geq |\rho_{14}|^2$ so that one of the conditions of Eq.~(\ref{eq:9}) is satisfied. Thus, provided also the other condition, namely $\rho_{11} > |\rho_{23}|^2$, is satisfied, an initially prepared entangled two-qubit state exhibits ESD and its negativity becomes zero at a finite time, say $t_1$.

Provided both conditions of Eq.~(\ref{eq:9}) are fulfilled, ESD can be delayed or even avoided by local unitary operations acting on the two qubits involved. In particular, let us concentrate on local unitary operations which exchange the density matrix elements
$\rho_{11}(t)$ and $\rho_{44}(t)$ of the quantum state at a time $t < t_1$ in such a way that their product, i.e. $\rho_{11}(t+0) \rho_{44}(t+0)$, remains constant but that the condition $\rho_{44}(t+0) > |\rho_{23}(t+0)|^2$ is violated. According to Eq.(\ref{eq:9}), in such a case ESD will be avoided. $\rho_{44}(t)$ is the probability of finding both qubits in their ground states. Thus, as a consequence of the dynamics of Eq.(\ref{eq:1}), the density matrix element $\rho_{44}(t)$ increases monotonically. There will be a limiting time $t_{sw}$ for any possible switching of these matrix elements for which ESD can still be avoided. If the local operation is applied after this limiting time, ESD may possibly be delayed but it is unavoidable. 

This simple consequence of the criterion of Eq.(\ref{eq:9}) explains recent numerical work on this problem \cite{RAA-Arx07}. In fact, operations of this type can avoid ESD for any initially prepared two-qubit X-state provided they are applied at a time $t < t_{sw}$. In particular, this applies to the subset of Werner states with $\rho_{14} = 0$ which are mixtures of a singlet state with probability $a$ and a completely unpolarized (chaotic) state.
These Werner states exhibit ESD in the parameter range $a \in [1/3, (-1 + \sqrt{5})/2)$ \cite{J-JPA06} where $\rho_{11} (t) > |\rho_{23}(t)|^2$ while entanglement decays asymptotically for values of $a$ in the range $ a \in ( (-1 + \sqrt{5})/2,1] $ which corresponds to $\rho_{11}(t) < |\rho_{23}(t)|^2 $.

Let us now deal with the question of which unitary transformations can achieve such a switch between $\rho_{44}(t)$ and $\rho_{11}(t)$. The most general $2 \times 2$ unitary matrix acting on a qubit is given by 
\begin{eqnarray}
U(2) = \left( \begin{array}{cc}
\cos(\theta) e^{i \alpha} & - \sin(\theta) e^{i (\alpha - \omega)} \\ 
\sin(\theta) e^{i(\beta + \omega)} & \cos(\theta) e^{i \beta} \end{array} \right),
\label{eq:12}
\end{eqnarray}
which is a linear superposition of the Pauli matrices $\sigma_x$, $\sigma_y$, $\sigma_z$, and the Identity matrix $\sigma_0$. Exchanging
the matrix elements $ \rho_{11}$ and $\rho_{44}$ can be achieved by applying two appropriately chosen unitaries $U_A$ and $U_B$ of the form of Eq.(\ref{eq:12}) on qubits $A$ and $B$ at a suitably chosen time, say $t$, so that the X-state $\rho_{X} (t)$ is transformed into another X-state $(U_A \otimes U_B) \rho_{X}(t) (U_A^{\dagger} \otimes U_B^{\dagger})$, for example. 

The most general local unitary operations transforming an arbitrary X-state into another one fulfill the conditions
\begin{eqnarray}
\sin(2 \theta_A) = \sin(2\theta_B) =  0 ~~\longrightarrow~~\nonumber \\ \theta_A = r_A \pi/2, \, \theta_B = r_B \pi/2, 
\label{eq:13}
\end{eqnarray}
with $r_A,r_B \in {\mathbb Z}$. X-state preserving local unitary transformations with even values of $r_A$ and $r_B$ do not have any significant effect on the density matrix elements except multiplying $\rho_{14}(t)$ by a constant phase factor. Odd values of $r_A$ and $r_B$ serve the purpose of exchanging $\rho_{11}(t)$ and $\rho_{44}(t)$. For any odd value of $r_A = r_B$, for example, the corresponding unitary two-qubit operator is given by
\begin{eqnarray}
U = \left( 
\begin{array}{cccc}
0 & 0 & 0 & e^{2 i (\alpha - \omega)} \\ 
0 & 0 & -e^{i (\alpha + \beta)} & 0 \\ 
0 & -e^{i (\alpha + \beta)} & 0 & 0 \\
e^{2 i (\beta + \omega)} & 0 & 0 & 0
\end{array} \right).
\label{eq:14}
\end{eqnarray}

A case in which such a X-state-preserving local unitary transformation is applied only onto qubit B can be described by parameters
$\theta_A = \alpha_A = \beta_A = \omega_A = 0$, for example. They lead to the transformations $\rho_{11}(t) \Leftrightarrow \rho_{22}(t)$, 
$\rho_{33}(t) \Leftrightarrow \rho_{44}(t)$, and $\rho_{14}(t) \Leftrightarrow \rho_{23}(t)$. In view of the characteristic time evolution of $\rho_{22}(t)$ in zero-temperature reservoirs (compare with Appendix A) and the criterion of Eq.(\ref{eq:9}) this implies that such a switch of matrix elements may delay ESD but it cannot be avoided.

{\bf Case 2:} For initially prepared entangled two-qubit X-states satisfying the alternative condition 
\begin{eqnarray}
\rho_{22} \rho_{33} &<& |\rho_{14}|^2, 
\label{eq:15}
\end{eqnarray}
the analytical expression for the negativity of the resulting quantum state $\rho_{X}(t)$ is given by 
\begin{eqnarray}
N_2 (\rho_{X}(t)) =&& max \, [\, 0, \, p \, (\sqrt{(\rho_{22} - \rho_{33})^2 + 4 |\rho_{14}|^2} \nonumber\\ && - (\rho_{22} + \rho_{33}) - (2 - 2 \, p) \, \rho_{11}) \, ].
\label{eq:16}
\end{eqnarray}
Eq.~(\ref{eq:15}) implies $\rho_{11} \rho_{44} \geq |\rho_{23}|^2$. Thus, the first condition of Eq.~(\ref{eq:9}) is always satisfied so that ESD occurs whenever also the second condition is satisfied. The simplest case arises for $\rho_{23} = 0$ where the initially prepared state is a Werner state, i.e. an incoherent mixture of a triplet state with probability $a$ and the completely  unpolarized state.
ESD takes place in the parameter regime $a \in [1/3, 1)$ where both conditions are satisfied during the time evolution. In the case of an initially prepared Bell state, i.e. for $a = 1$, the second condition of Eq.~(\ref{eq:9}) fails and entanglement decays asymptotically.

As discussed above, the first condition of Eq.~(\ref{eq:9}) is always fulfilled in the cases considered here so that ESD takes place always except in the particular case of an initially prepared Bell state which fulfills the condition $(\rho_{11}(t) + \rho_{22}(t))(\rho_{11}(t) + \rho_{33}(t)) = |\rho_{14}(t)|^2$. As a consequence any switch capable of exchanging $ \rho_{23}(t) $ and $\rho_{14}(t)$ will be sufficient to avoid or delay ESD. Such a switch can be implemented by a local unitary X-state-preserving transformation acting on qubit $A$ or $B$ only. As a result the second condition of Eq.~(\ref{eq:9}) remains always true, while the validity of the first condition depends on the choice of the switching time $t$. If $\rho_{33}(t) > |\rho_{14}(t)|^2$ ESD is unavoidable.
However, for all switching times violating this condition ESD is averted completely. As an example, let us consider Werner states with $\rho_{23}(0) = 0$. In the parameter range $a \in [1/3, (-1 + \sqrt{5})/2)$ the condition $\rho_{33}(t) > |\rho_{14}(t)|^2$ is fulfilled for all times so that ESD takes place. In the parameter regime $a \in ((-1 + \sqrt{5})/2, 1]$ this condition is violated so that entanglement decays asymptotically.

\subsection{Controlling ESD in statistically independent finite-temperature reservoirs}

According to Eq.(\ref{eq:6}), ESD takes place if and only if the two principal minors 
\begin{eqnarray}
[\rho^{PT}(14)] = m^2 (m + 1)^2 + e^{-(2 m + 1) \gamma t} [F^{(1)}] \nonumber \\ + e^{-2 (2 m + 1) \gamma t} [F^{(2)}] + e^{-3 (2 m + 1) \gamma t} [F^{(3)}] \nonumber \\ + e^{-4 (2 m + 1) \gamma t} [F^{(4)}],\nonumber
\end{eqnarray}

\begin{eqnarray}
[\rho^{PT}(23)] = m^2 (m + 1)^2 + e^{-(2 m + 1) \gamma t} [G^{(1)}] \nonumber \\ + e^{-2 (2 m + 1) \gamma t} [G^{(2)}] + e^{-3 (2 m + 1) \gamma t} [G^{(3)}] \nonumber \\ + e^{-4 (2 m + 1) \gamma t} [G^{(4)}],
\label{eq:17}
\end{eqnarray}
are positive in the limit of very long interaction times. For simplicity, we have taken $ m = n$ and $\gamma_1 = \gamma_2 = \gamma$ in Eq.(\ref{eq:17}). The quantities $F^{(i)}$ and $G^{(i)}$ are functions of $m$ and of the initial matrix elements of the initially prepared quantum state. Their explicit forms are given in Appendix C. Furthermore, the general expressions for $[\rho^{PT}(14)]$ and $[\rho^{PT}(23)]$ are provided in Appendix B. For sufficiently long times, say $t\geq t_{\infty}^-$ and for $m > 0$, factors of the form $e^{-(2 m + 1) \gamma t_{\infty}^-}$ are exponentially small and therefore both $[\rho^{PT}(14)]$ and $[\rho^{PT}(23)]$ are positive. Analogously, one can show that for unequal values of the mean photon numbers $m$ and $n$, both minors are positive if and only if at least one of these mean photon numbers is not equal to zero. Hence, we arrive at the central result that if one of the (photon) reservoirs is at nonzero temperature all initially prepared X-states exhibit ESD.

As ESD is unavoidable in these cases, it may be useful at least to delay it. Indeed this can be achieved for all possible X-states \cite{YE-PRL04,RAA-Arx07}. For the sake of demonstration let us consider the particular example of an initially prepared entangled state of the form 
\begin{eqnarray}
\rho = \frac{1}{3} ( |1,1\rangle\langle1,1| + 2 |\Psi\rangle\langle\Psi|)
\label{eq:18}
\end{eqnarray}
with $|\Psi\rangle = (|0,1\rangle \pm |1,0\rangle)/\sqrt{2}$. It is known that this entangled state, while interacting with a vacuum reservoir, looses its entanglement at $\gamma t \approx 0.5348$ \cite{RAA-Arx07}. However, while interacting with reservoirs at finite temperatures, the time of sudden death for this initial state depends on the values of $m$  and $n$. The solution of Eq.~(\ref{eq:1}) for the input state of Eq.~(\ref{eq:18}) can be obtained easily using the general solution given by Eq.~(\ref{A1}). After taking the partial transpose of the resulting density operator it is possible to obtain an analytical expression for the negativity of the 
quantum state at any time $t$. Setting $ m = n = 0.1$, for example, we observe that ESD occurs at time $t_{ESD} \approx 0.4115/\gamma$. 
Depending on the time when local unitary transformations are applied to qubits $A$ and $B$, ESD can be speeded up or delayed for some finite time. 
\begin{figure}[h]
\scalebox{2.0}{\includegraphics[width=1.7in]{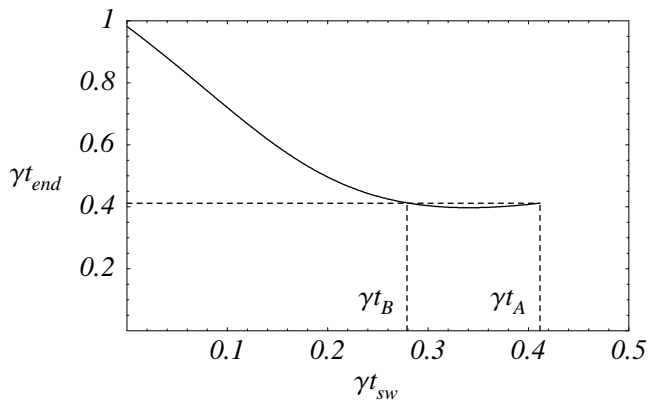}}
\caption{Dependence of
the time $t_{end}$ of ESD on the switching time $t_{sw}$ for $m = n = 0.1$:
The X-state-preserving local unitary transformations switch the density matrix elements $\rho_{11}$ and $\rho_{44}$ in Eq.~(\ref{eq:18}).
Starting on the right at switching time $t_A \approx 0.4115/\gamma$
this dependence exhibits a broad and small dip before rising to the maximum possible time $t_{end} \approx 0.9817/\gamma$.}
\label{FIG:2}
\end{figure} 

Fig. (\ref{FIG:2}) displays the time $t_{end}$ at which ESD takes place and its dependence on the time of switching $t_{sw}$.
The earlier appropriate local unitary transformations are applied, the more ESD is delayed. However, typically such a delay is possible only for a certain range of switching times $t_{sw}$, such as $t_{sw} < t_B = 0.279/\gamma$ in Fig.(\ref{FIG:2}). Eventually ESD is unavoidable. In the case considered in Fig.(\ref{FIG:2}), it takes place at $t_{end} \approx 0.9817/\gamma$. 
\begin{figure}[h]
\scalebox{2.0}{\includegraphics[width=1.7in]{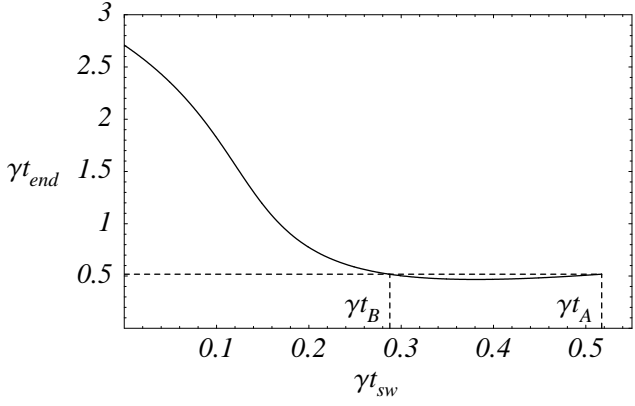}}
\caption{Dependence of the time $t_{end}$ of ESD on the switching time $t_{sw}$ for $m = n = 0.01$:
The local unitary transformation are the same as in Fig.(\ref{FIG:2}).
For switching times below $t_A \approx 0.5172/\gamma$
this dependence has a broad and small dip before rising to the maximum possible time $t_{end} \approx 2.7087/\gamma$.}
\label{FIG:3}
\end{figure}

In Fig.(\ref{FIG:3}) the relation between $t_{end}$ and $t_{sw}$ is depicted for mean thermal photon numbers $m = n = 0.01$. In this case ESD occurs at $t_{ESD} \approx 0.5172/\gamma$. If the switch is applied before $t_{A} \approx 0.5172/\gamma$, ESD is hastened. Any switch made before $t_{B} \approx 0.2877/\gamma$ delays ESD  up to the maximum possible time $t_{end} \approx 2.7087/\gamma$. This larger delay in comparison with the case considered in Fig.(\ref{FIG:2}) is due to smaller values of the mean photon numbers.

Two recent papers that have appeared since completion of our work examine related topics. One studies two harmonic oscillators coupled to a common environment also modelled as oscillators \cite{PAZ08}. ESD and entanglement revival are examined as well as non-Markovian environments. The other studies two interacting qubits in a magnetic field and a thermal Markovian environment, presenting nonmonotonic relaxation rates as functions of the magnetic field and temperature \cite{DUBI08}. 

\section{Conclusions}

A criterion has been presented characterizing the conditions which lead to ESD of X-states of two qubits coupled to statistically independent reservoirs at finite temperatures. Based on this criterion, we have presented an analytical description of ESD of X-states and its delaying or its avoidance by local unitary actions. We have proved that if at least one of the reservoirs is at finite temperature, all X-states exhibit ESD. Thus, in these cases ESD can only be delayed but not averted. Preliminary studies \cite{ARA-UP08} indicate that similar results also hold for qubit-qutrit systems. 

\begin{acknowledgments}
M. Ali acknowledges financial support by the Higher Education Commission, Pakistan, and by the DAAD.
\end{acknowledgments}

\appendix

\section{}

Solutions of Eq.(\ref{eq:1}) for some special initially prepared X-states have been provided in the Appendix of Ref.\cite{ILZ-PRA07}. In this Appendix we provide the most general solutions of Eq.(\ref{eq:1}) for any initially prepared two-qubit quantum state $\rho$ with density matrix elements $\rho_{ij}$ in the computational basis introduced in Sec. II. These solutions are given by
\begin{eqnarray*}
\rho_{11}(t) &=& \frac{1}{(2 m + 1)(2 n + 1)} \{ m n \, + m \, [\, (n + 1) \rho_{11} + \rho_{33} \\ && - n (\rho_{22} - \rho_{33} + \rho_{44}\, ) \, ] e^{-(2 n + 1) \gamma_2 t} + n [ \, (m +1) \\ && \rho_{11} + (m + 1) \rho_{22} - m ( \rho_{33} + \rho_{44} ) \, ]  e^{-(2 m + 1) \gamma_1 t}  \\ && + [ \, (m+1) (n+1) \rho_{11} - m \rho_{33} - n (\rho_{22} + m \rho_{22} \\ && + m \rho_{33} - m \rho_{44} \, ) \, ] e^{- [(2 m +1) \gamma_1 + (2 n + 1) \gamma_2 ] t} \},
\end{eqnarray*}

\begin{eqnarray*}
\rho_{22}(t) &=& \frac{1}{(2 m + 1)(2n+1)} \{ m (n + 1) - m [ (n+1) \rho_{11} \\ && + \rho_{33} - n (\rho_{22} - \rho_{33} +  \rho_{44}) ] e^{-(2n+1) \gamma_2 t} + \\ && (n+1) [ \, (1+m) \, \rho_{11} + (m+1) \, \rho_{22} - m \\ && \times ( \rho_{33} + \rho_{44}\, ) \, ] \, e^{-(2 m + 1) \gamma_1 t} + [-(m+1) \\ && \times (n+1) \, \rho_{11} + m \, \rho_{33} + n \, ( \, (m+1) \, \rho_{22} + \\ && m \, \rho_{33}  - m \, \rho_{44} \, ) \, ] \, e^{-[ (2 m + 1) \gamma_1 + (2 n + 1)\gamma_2] t} \},
\end{eqnarray*}

\begin{eqnarray*}
\rho_{33}(t) &=& \frac{1}{(2 m + 1)(2n+1)} \{ n (m + 1) + (m+1) \\ && \times [ \, (n+1) \, \rho_{11} + \rho_{33} - n (\rho_{22} - \rho_{33} + \rho_{44} \, ) \, ] \\ && \times e^{-(2 n + 1) \gamma_2 t} - n \, [ \, (m+1) \, \rho_{11} + (m+1) \, \rho_{22} \\ && - m \, (\rho_{33} + \rho_{44} \, ) \, ] \, e^{-(2m+1)\gamma_1 t} + [ - (m+1) \\ && \times (n+1) \, \rho_{11} + m \, \rho_{33} + n \, ( \, (m+1) \, \rho_{22} \\ && + m \, \rho_{33} - m \, \rho_{44} \, ) \, ] \, e^{-[(2 m + 1) \gamma_1 + (2n+1)\gamma_2] t} \},
\end{eqnarray*}


\begin{eqnarray*}
\rho_{12}(t) &=& \frac{1}{2m+1} \{ \, m (\rho_{12} + \rho_{34}) e^{\frac{-1}{2} (2n + 1) \gamma_2 t} + [ (m + 1) \\ && \times \rho_{12} - m \rho_{34} ] \, e^{\frac{-1}{2} [2 (2m+1) \gamma_1 + (2n+1) \gamma_2]t} \, \},
\end{eqnarray*}

\begin{eqnarray*}
\rho_{13}(t) &=& \frac{1}{2n+1} \{ \, n (\rho_{13} + \rho_{24}) e^{\frac{-1}{2} (2m+1) \gamma_1 t} + [ (n + 1) \\ && \times \rho_{13}  - n \rho_{24}] \, e^{\frac{-1}{2}[(2m+1) \gamma_1 + 2  (2n+1) \gamma_2]t} \, \},
\end{eqnarray*}

\begin{eqnarray*}
\rho_{24}(t) &=&  \frac{1}{2n+1} \{ \, (n + 1) (\rho_{13} + \rho_{24}) e^{\frac{-1}{2} (2m + 1) \gamma_1 t} + \\ && [n \rho_{24} - (n + 1) \rho_{13}] \, e^{\frac{-1}{2}[(2m+1)\gamma_1 + 2 (2n+1)\gamma_2]t} \, \},
\end{eqnarray*}

\begin{eqnarray*}
\rho_{34}(t) &=& \frac{1}{2m+1} \{ \, (m + 1) (\rho_{12} + \rho_{34}) e^{\frac{-1}{2} (2n+1) \gamma_2 t} + \\ && [m \rho_{34} - (m + 1) \rho_{12} ] \, e^{\frac{-1}{2}[2(2m+1)\gamma_1 + (2n+1)\gamma_2]t} \, \},
\end{eqnarray*}

\begin{eqnarray*}
\rho_{14}(t) &=& \rho_{14} \, e^{-[(m + \frac{1}{2}) \gamma_1 + (n + \frac{1}{2})\gamma_2] t},
\end{eqnarray*}

\begin{eqnarray}
\rho_{23}(t) &=& \rho_{23} \, e^{-[(m + \frac{1}{2}) \gamma_1 + (n + \frac{1}{2})\gamma_2] t}.
\label{A1}
\end{eqnarray}

\section{}

For X-states, i.e. quantum states with
$\rho_{12}=\rho_{13}=\rho_{24}=\rho_{34}=0$, the dependence of all seven principal-minors of Eq.~(\ref{eq:2}) on 
$[ \rho^{PT}(14)]$ and on $[ \rho^{PT}(23)]$ is given by

\begin{eqnarray*}
[\rho^{PT}(14)] &=& \left| \begin{array}{cc}
                 \rho_{11} & \rho_{23} \\
                 \rho_{32} & \rho_{44}
                  \end{array} \right| = \rho_{11} \, \rho_{44} - |\rho_{23}|^2 \, ,
\end{eqnarray*}

\begin{eqnarray*}
[\rho^{PT}(23)] &=& \left| \begin{array}{cc}
                 \rho_{22} & \rho_{14} \\
                 \rho_{41} & \rho_{33}
                  \end{array} \right| = \rho_{22} \, \rho_{33} - |\rho_{14}|^2 \, ,
\end{eqnarray*}

\begin{eqnarray*}
[\rho^{PT}(123)] &=& \left| \begin{array}{ccc}
                 \rho_{11} & 0 & 0 \\
                 0 & \rho_{22} & \rho_{14} \\
                 0 & \rho_{41} & \rho_{33}
                  \end{array} \right| = \rho_{11} \, [\rho^{PT}(23)] \, ,
\end{eqnarray*}

\begin{eqnarray*}
[\rho^{PT}(124)] &=& \left| \begin{array}{ccc}
                 \rho_{11} & 0 & \rho_{23} \\
                 0 & \rho_{22} & 0 \\ 
                 \rho_{32} & 0 & \rho_{44}
                  \end{array} \right| = \rho_{22} \, [\rho^{PT}(14)] \, ,
\end{eqnarray*}

\begin{eqnarray*}
[\rho^{PT}(134)] &=& \left| \begin{array}{ccc}
                 \rho_{11} & 0 & \rho_{23} \\
                  0 & \rho_{33} & 0 \\ 
                 \rho_{32} & 0 & \rho_{44}
                  \end{array} \right| = \rho_{33} \, [\rho^{PT}(14)] \,,
\end{eqnarray*}

\begin{eqnarray*}
[\rho^{PT}(234)] &=& \left| \begin{array}{ccc}
                 \rho_{22} & \rho_{14} & 0 \\
                 \rho_{41} & \rho_{33} & 0 \\ 
                  0 & 0 & \rho_{44}
                  \end{array} \right| = \rho_{44} \, [\rho^{PT}(23)] \,,
\end{eqnarray*}

\begin{eqnarray}
[\rho^{PT}(1234)] =&& \left| \begin{array}{cccc}
                 \rho_{11} & 0 & 0 & \rho_{23} \\
                  0 & \rho_{22} & \rho_{14} & 0 \\ 
                  0 & \rho_{41} & \rho_{33} & 0 \\
                 \rho_{32} & 0 & 0 & \rho_{44}
                  \end{array} \right| \nonumber \\
=&& [\rho^{PT}(14)] \, [\rho^{PT}(23)] \,.
\label{B1}
\end{eqnarray}

\section{}

The expressions for $F^{(i)}$ in Eq.~(\ref{eq:17}) are given by
\begin{eqnarray*}
F^{(1)} &=& m \, (m + 1) \, [\, (2 m + 1) \, \rho_{11} + \rho_{22} + \rho_{33} - 2 \, m \, \rho_{44} \, ],
\end{eqnarray*}

\begin{eqnarray*}
F^{(2)} &=& -2 \, m^4 \, [2 \, \rho_{44}^2 - \rho_{44} + \rho_{22} + 8 \, |\rho_{23}|^2 + \rho_{33} \, ] + \\ && 2 \, m^3 \, [-2 \, \rho_{44}^2 + 2 \, \rho_{33} \, \rho_{44} + \rho_{44} - 16 \, |\rho_{23}|^2 - 2 \, \rho_{33} \\ && + 2 \, \rho_{22} \, (\rho_{44} - 1 ) \, ] - m^2 \, [ \, \rho_{22}^2 + (2 \, \rho_{33} - 4 \, \rho_{44} \\ && + 3 ) \, \rho_{22} + \rho_{33}^2 + 24 \, |\rho_{23}|^2 + 3 \, \rho_{33} - 4 \, \rho_{33} \, \rho_{44} \\ && - \rho_{44}\, ] - 4 \, m \, (m +1)^3 \, \rho_{11}^2 - m \, [\rho_{22}^2 + 2 \, \rho_{33} \, \rho_{22} \\ && + \rho_{22} + \rho_{33}^2 + 8 \, |\rho_{23}|^2 + \rho_{33} \, ] - |\rho_{23}|^2 + (m +1)^2 \\ && \rho_{11} [\, (8 \, \rho_{44} + 2 ) \, m^2 + (-4 \rho_{22} - 4 \rho_{33} + 2 ) m + 1 \, ],
\end{eqnarray*}

\begin{eqnarray*}
F^{(3)} &=& -2 \, \rho_{11}^2 \, (m+1)^3 + (m + 1) \, \rho_{11} \, [(2 m^2 + m - 1) \\ && \times (\rho_{22} + \rho_{33}) + 2 \, m \, \rho_{44} \, ] \, + m \, [\, (m + 1) \, \rho_{22}^2 \\ && + (2 (m + 1) \, \rho_{33} - m \, (2 m + 3 ) \, \rho_{44}) \, \rho_{22} + \\ && (m + 1) \, \rho_{33}^2 + 2 \, m^2 \, \rho_{44}^2 - m \, (2 m + 3) \, \rho_{33} \, \rho_{44}\, ],
\end{eqnarray*}

\begin{eqnarray}
F^{(4)} =&& \, [ \, m^2 \, (\rho_{11} - \rho_{22} - \rho_{33} + \rho_{44}) + \nonumber \\ 
&& m \, (2 \, \rho_{11} - \rho_{22} - \rho_{33}) + \rho_{11}]^2.
\label{C1}
\end{eqnarray}

Similarly, the expressions for $G^{(i)}$ in Eq.~(\ref{eq:17}) are given by
\begin{eqnarray*}
G^{(1)} &=& m \, (m + 1) \, [\, (2 m + 1) \, \rho_{11} + \rho_{22} + \rho_{33} - 2 \, m \, \rho_{44} \, ],
\end{eqnarray*}

\begin{eqnarray*}
G^{(2)} &=& -2 m^4 \, [\, 2 \rho_{22}^2 - 4 \rho_{33} \, \rho_{22} - \rho_{22} + 2 \rho_{33}^2 + \rho_{11} - \rho_{33} \\ && + 8 |\rho_{14}|^2 + \rho_{44} \, ] - 2 m^3 \, [ \, 4 \rho_{22}^2 - 8 \rho_{33} \, \rho_{22} - 2 \rho_{22} \\ && + 4 \rho_{33}^2 + 3 \rho_{11} - 2 \rho_{33} + 16 |\rho_{14}|^2 + \rho_{44} \, ] + m^2 \\ && [ \rho_{11}^2 - 2 (\rho_{44} + 3 ) \, \rho_{11} - 6 \rho_{22}^2 - 6 \rho_{33}^2 + \rho_{44}^2 + 2 \rho_{22} \\ && + 12 \, \rho_{22} \, \rho_{33} + 2 \, \rho_{33} - 24 \, |\rho_{14}|^2 \, ] + m \, [2 \, \rho_{11}^2 + (\rho_{22} \\ && + \rho_{33} - 2 \, \rho_{44} - 2 ) \, \rho_{11} - 2 \, \rho_{22}^2 - 2 \, \rho_{33}^2 + 4 \, \rho_{22} \, \rho_{33} \\ && - 8 |\rho_{14}|^2 - \rho_{22} \, \rho_{44} - \rho_{33} \, \rho_{44} \, ] + \rho_{11}^2 + \rho_{22} \, \rho_{33} \\ && + \rho_{11} \, (\rho_{22} + \rho_{33}) - |\rho_{14}|^2,
\end{eqnarray*}

\begin{eqnarray*}
G^{(3)} &=& -2 \, \rho_{11}^2 \, (m+1)^3 + (m + 1) \, \rho_{11} \, [ \, (2 \, m^2 + m - 1) \\ && \times (\rho_{22} + \rho_{33}) + 2 \, m \, \rho_{44} \, ] + m \, [ \, (m + 1) \, \rho_{22}^2 + ( \, 2 \\ && (m + 1) \, \rho_{33} - m \, (2 m + 3) \, \rho_{44} \, ) \, \rho_{22} + ( m + 1) \,  \rho_{33}^2 \\ && + 2 \, m^2 \, \rho_{44}^2 - m \, (2 m + 3) \, \rho_{33} \, \rho_{44} \, ],
\end{eqnarray*}

\begin{eqnarray}
G^{(4)} =&& [m^2 (\rho_{11} - \rho_{22} - \rho_{33} + \rho_{44}) + \nonumber \\
&& m (2 \rho_{11} - \rho_{22} - \rho_{33}) + \rho_{11}]^2.
\label{C2}
\end{eqnarray}


\begin{thebibliography}{}

\bibitem{Nielsen2000} M. A. Nielsen and I. L. Chuang, {\it Quantum Computation and Quantum Information} (Cambridge Univ. Press, Cambridge, 2000).

\bibitem{E-PRL91} A. K. Ekert, Phys. Rev. Lett. {\bf 67}, 661 (1991).

\bibitem{BBCJPW-PRL93} C. H. Bennett, G. Brassard, C. Crepeau, R. Josza, A. Peres, and W. K. Wootters, Phys. Rev. Lett. {\bf 70}, 1895 (1993); Dik Bouwmeester, J. W. Pan, K. Mattle, M. Eibl, H. Weinfurter, and A. Zeilinger, Nature {\bf 390}, 575 (1997).

\bibitem{YE-PRB03} T. Yu and J. H. Eberly, Phys. Rev. B {\bf 66}, 193306 (2002); T. Yu and J. H. Eberly, Phys. Rev. B {\bf 68}, 165322 (2003).

\bibitem{YE-PRL04} T. Yu and J. H. Eberly, Phys. Rev. Lett. {\bf 93}, 140404 (2004).

\bibitem{YE-OC06} T. Yu and J. H. Eberly, Opt.Commun. {\bf 264}, 393 (2006).

\bibitem{YE-PRL06} T. Yu and J. H. Eberly, Phys. Rev. Lett. {\bf 97}, 140403 (2006).

\bibitem{EY-Sc07} J. H. Eberly and T. Yu, Science {\bf 316}, 555 (2007).

\bibitem{AM-Sc07} M. P. Almeida, F. de Melo, M. Hor-Meyll, A. Salles, S. P. Walborn, P. H. S. Ribeiro, and L. Davidovich, Science {\bf 316}, 579 (2007).

\bibitem{LC-PRL07} J. Laurat, K. S. Choi, H. Deng, C. W. Chou, and H. J. Kimble, Phys. Rev. Lett. {\bf 99}, 180504 (2007).

\bibitem{LR-PRA07} F. Lastra, G. Romero, C. E. Lopez, M. Fran\c ca Santos and J. C. Retamal, Phys. Rev A {\bf 75}, 062324 (2007).

\bibitem{SW-PRA07} Z. Sun, X. Wang and C. P. Sun, Phys. Rev A {\bf 75}, 062312 (2007).

\bibitem{AQJ-PRA08} A. Al-Qasimi and D. F. V. James, Phys. Rev. A {\bf 77}, 012117 (2008).

\bibitem{ILZ-PRA07} M. Ikram, Fu-li Li, and M. S. Zubairy, Phys. Rev. A {\bf 75}, 062336 (2007).

\bibitem{AJ-PLA07} K. Ann and G. Jaeger, Phys. Lett. A {\bf 372}, 579 (2008).

\bibitem{ARR-Arx07} M. Ali, A. R. P. Rau and K. Ranade, arXiv:quant-ph/0710.2238.

\bibitem{AJ-PRA07} K. Ann and G. Jaeger, Phys. Rev. A {\bf 76} 044101 (2007).

\bibitem{CZ-PRA08} X. Cao and H. Zheng, Phys. Rev. A {\bf 77}, 022320 (2008).

\bibitem{DJ-PRA06} \L. Derkacz and L. Jak\'obczyk, Phys. Rev. A {\bf 74}, 032313 (2006).

\bibitem{J-JPA06} A. Jamr\'oz, J. Phys. A {\bf 39}, 7727 (2006).

\bibitem{YE-QIQC07} T. Yu and J. H. Eberly, {\it Quantum Information and Computation} {\bf 7}, 459 (2007); M. Y\"ona\c c, T. Yu and J. H. Eberly, J. Phys. B {\bf 39}, S621 (2006).

\bibitem{RAA-Arx07} A. R. P. Rau, M. Ali and G. Alber, EPL {\bf 82}, 40002 (2008).

\bibitem{ARA-UP08} M. Ali, A. R. P. Rau and G. Alber, in preparation.

\bibitem{Pe-PRL96} A. Peres, Phys. Rev. Lett. {\bf 77}, 1413 (1996).

\bibitem{HHH-PLA96} M. Horodecki, P. Horodecki, and R. Horodecki, Phys. Lett. A {\bf 223}, 1 (1996); G. Alber {\it et al} {\it Quantum Information} (Springer-Verlag, Berlin, 2001), p.151.

\bibitem{HZ-PRA07} Jie-Hui Huang and Shi-Yao Zhu, Phys. Rev. A {\bf 76}, 062322 (2007).

\bibitem{Scully-Zubairy1997} M. O. Scully and M. S. Zubairy, {\it Quantum Optics} (Cambridge University Press, London, 1997).

\bibitem{VW-PRA02} G. Vidal and R. F. Werner, Phys. Rev. A {\bf 65}, 032314 (2002).

\bibitem{STV-PRA98} A. Sanpera, R. Tarrach and G. Vidal, Phys. Rev. A {\bf 58}, 826 (1998).

\bibitem{HJ-1985} R. A. Horn and C. R. Johnson, {\it Matrix Analysis} (Cambridge University Press, Cambridge, 1985), Chap. 7.

\bibitem{Wer-PRA89} R. F. Werner, Phys. Rev. A {\bf 40}, 4277 (1989).

\bibitem{JJ-PLA04} L. Jak\'obczyk and A. Jamr\'oz, Phys. Lett. A {\bf 333}, 35 (2004).

\bibitem{SMD-PRA06} M. Fran\c ca Santos, P. Milman, L. Davidovich, and N. Zagury, Phys. Rev. A {\bf 73}, 040305(R) (2006).

\bibitem{PAZ08} J. P. Paz and A. J. Roncaglia. Phys. Rev. Lett. {\bf 100}, 220401 (2008).

\bibitem{DUBI08} Y. Dubi and M. Di Ventra, arXiv:quant-ph/0809.3256.

\end{thebibliography}
\end{document}